\documentclass{svproc}
\usepackage{url}

\usepackage{graphicx}
\usepackage{float}
\usepackage{xcolor}
\usepackage{soul}
\usepackage{subcaption}

\usepackage{fancyhdr}

\begin{document}
\mainmatter              
\title{Impact of Edge Software Evolution on Audio Tagging Models: An Empirical Study on NVIDIA Jetson Orin Nano}
\titlerunning{Impact of Edge Software Evolution on Audio Tagging}  
%
\author{Guillem Fustier-Castillo \and Ruben Ribes-Serrano \and Jordi Grau-Haro \and Marta Garcia-Ballesteros \and Javier Naranjo-Alcazar
}
\authorrunning{Guillem Fustier et al.} 
%
\tocauthor{}
\institute{Instituto Tecnologico de Informatica, Valencia, Spain,\\
\email{\{gfustierbcr,jgrau,rribes,martagarcia,jnaranjo\}@iti.es}
}

\maketitle              

\begin{abstract}

Audio tagging is a fundamental task in machine listening, enabling the automatic identification of acoustic events in applications such as intelligent monitoring, industrial inspection, environmental sensing, and smart environments. As these applications increasingly migrate to edge platforms, deployment-oriented metrics such as latency, energy consumption, and resource utilization become critical alongside predictive performance. However, most existing studies evaluate audio tagging models under a single software environment, overlooking the potential impact of software stack evolution on deployment efficiency. This paper investigates how software ecosystem updates affect the execution characteristics of state-of-the-art audio tagging architectures on edge devices. We evaluate representative convolutional and Transformer-based models, including PANNs (CNN and ResNet variants), AST, and PaSST, on an NVIDIA Jetson Orin Nano platform using two generations of NVIDIA's software ecosystem: JetPack 6.2 and JetPack 7.2. Under identical hardware conditions, we analyze inference latency, energy consumption, CPU/GPU utilization, and memory footprint across multiple inference backends and power modes. The results show that software stack evolution can substantially affect deployment performance. JetPack 7.2 reduces latency and energy consumption by more than 50\% for several convolutional architectures when combined with TensorRT, while significantly increasing GPU utilization. In contrast, AST exhibits only marginal gains, whereas PaSST shows intermediate behavior. These findings demonstrate that software evolution is a critical factor in Edge AI evaluation and can alter the relative deployment efficiency of competing architectures. Consequently, software stack versions should be explicitly considered when benchmarking and selecting audio tagging models for embedded deployment.

\end{abstract}
\section{Introduction}

Recent advances in deep learning have significantly transformed the field of machine listening, enabling machines to automatically analyze, interpret, and understand acoustic environments \cite{mesaros2016tut,jiang2026multi,ciaburro2026sound,niu2026acavcaps}. Machine listening encompasses a broad range of tasks, including acoustic scene classification \cite{naranjo2020acoustic}, sound event detection \cite{mesaros2021sound}, anomalous sound detection \cite{suefusa2020anomalous}, and audio tagging \cite{gong2021psla}. Among these, audio tagging has emerged as one of the most relevant challenges due to its applicability in intelligent monitoring systems, smart cities \cite{nogueira2022sound}, environmental sensing \cite{gairi2025environmental} and surveillance \cite{nogueira2022sound,10.1145/2871183},  autonomous robots \cite{lin2024advancing}, among others.

Audio tagging consists of assigning one or more semantic labels to an audio recording according to the acoustic events present in the signal \cite{schmid2023efficient}. Unlike sound event detection, which requires temporal localization of events, audio tagging focuses on clip-level classification, making it an attractive solution for large-scale acoustic perception systems. The availability of large annotated datasets, particularly AudioSet \cite{Gemmeke:2017:ICASSP}, has accelerated progress in this field and has enabled the development of increasingly powerful deep neural network architectures.

The last decade has witnessed a rapid evolution of audio tagging models. Convolutional Neural Networks (CNNs) initially dominated the field due to their ability to efficiently learn hierarchical time-frequency representations from spectrograms \cite{xu2017convolutional,grau2026comprehensive}. A notable milestone was the introduction of Pretrained Audio Neural Networks (PANNs) \cite{Kong:2020:TASLP}, which demonstrated that large-scale supervised pretraining on AudioSet could significantly improve transferability across a wide range of audio understanding tasks. PANNs established a strong CNN-based baseline and remain one of the most widely adopted architectures for audio tagging and audio representation learning \cite{zhang2026few,liu2022leveraging}.

More recently, the success of Transformer architectures in natural language processing \cite{zhang2024survey} and computer vision \cite{han2022survey} has inspired their adoption in machine listening. The Audio Spectrogram Transformer \cite{gong2021ast} (AST) introduced a purely attention-based architecture for audio classification, eliminating convolutional layers and achieving state-of-the-art performance on several benchmark datasets. AST demonstrated that self-attention mechanisms can effectively model long-range temporal dependencies and global acoustic context, challenging the long-standing dominance of CNN-based approaches.

Following AST, several Transformer-based architectures have been proposed to improve computational efficiency while preserving accuracy. Among them, the Patchout Audio Spectrogram Transformer \cite{koutini2021efficient} (PaSST) introduced a patch-dropping strategy that reduces computational complexity during training and inference while maintaining competitive performance. PaSST has become a representative example of the current generation of efficient audio Transformers.

While research has traditionally focused on improving recognition performance \cite{Kong:2020:TASLP,gong2021psla,singh2024atgnn}, practical deployment requirements are becoming increasingly important \cite{gairi2025environmental,bibbo2023audio}. Many machine listening applications operate under strict latency, power consumption, memory, and connectivity constraints \cite{naranjo2025real}. Examples include industrial monitoring systems \cite{antonini2018smart}, autonomous robots, smart sensors \cite{lamrini2023evaluating}, wildlife monitoring platforms \cite{schwartz2021deployment}, and intelligent surveillance systems \cite{wang2025transferable}. In such scenarios, transmitting raw audio streams to remote cloud infrastructures may introduce unacceptable latency, increase communication costs, consume excessive bandwidth, and raise privacy concerns \cite{al2024computing,trigkas2026federated}.

These limitations have motivated the emergence of Edge AI, where machine learning models are deployed directly on devices located close to the data source \cite{pandya2026acoustimeal}. By moving inference from centralized cloud infrastructures to edge devices, systems can provide faster responses, improved privacy, reduced network dependence, and lower operational costs. Consequently, Edge AI has become a key enabling technology for real-time intelligent systems.

Deploying deep neural networks at the edge, however, remains challenging. Embedded platforms are constrained by limited computational resources, memory capacity, thermal envelopes, and energy budgets. As a result, evaluating AI models solely according to predictive accuracy is insufficient for real-world deployment. Metrics such as inference latency, energy consumption, and resource utilization become equally important when assessing model suitability for edge environments. Recent surveys on embedded audio intelligence emphasize that resource efficiency is one of the primary obstacles preventing the widespread adoption of state-of-the-art audio models in practical systems \cite{gairi2025environmental,vandendriessche2021environmental}.

The NVIDIA Jetson family has emerged as one of the most widely used hardware platforms for Edge AI deployment \cite{mittal2019survey}. These devices provide heterogeneous computing resources that combine ARM CPUs, CUDA-enabled GPUs, Tensor Cores, and specialized acceleration libraries. In practice, the performance achieved by a deployed model depends not only on its architecture but also on the underlying software stack, including operating system components, CUDA libraries, TensorRT optimizations, compiler toolchains, and runtime frameworks.


To the best of our knowledge, limited attention has been devoted to understanding how software stack evolution affects the deployment characteristics of state-of-the-art audio tagging models on edge devices \cite{grau2026comprehensive}. Existing literature primarily focuses on architectural comparisons, model compression techniques, quantization strategies, or hardware benchmarking \cite{zhang2026review,belatik2025optimization}. The potential impact of software ecosystem updates on latency, energy efficiency, and resource utilization remains largely unexplored \cite{khan2026edge}.

This paper addresses this gap through an empirical study of three representative state-of-the-art audio tagging architectures: PANNs (both CNNs and ResNet based), AST, and PaSST. We evaluate their deployment on an NVIDIA Jetson Orin Nano platform under two generations of the NVIDIA software ecosystem, JetPack 6.2 and JetPack 7.2. Our analysis investigates how software evolution influences inference latency, energy consumption, and resource utilization across different architectural paradigms, including convolutional and Transformer-based models.

The contributions of this work are fourfold:

\begin{itemize}
    \item We provide a systematic edge deployment evaluation of three state-of-the-art audio tagging architectures representing distinct design philosophies.
    \item We quantify the impact of software stack evolution on inference performance and energy efficiency using the same hardware platform.
    \item We demonstrate that software updates do not affect all architectures equally, revealing substantial differences in deployment gains across model families.
    \item We discuss the implications of these findings for Edge AI benchmarking and model selection, showing that deployment conclusions may change significantly across software generations.
\end{itemize}

The results suggest that software evolution should be considered a key factor in Edge AI evaluation and that architecture selection for embedded machine listening applications cannot be decoupled from the software ecosystem in which models are deployed.

\section{Related work}\label{sec:related}

\subsection{Machine Listening and Audio Tagging}\label{subsec:audio_taggig}

Machine listening has experienced rapid progress in recent years, largely driven by the availability of large-scale datasets such as AudioSet \cite{Gemmeke:2017:ICASSP} and the adoption of deep learning techniques for audio understanding tasks. Within audio tagging, convolutional neural networks (CNNs) have long represented the dominant architectural paradigm, with Pretrained Audio Neural Networks (PANNs) \cite{Kong:2020:TASLP} becoming one of the most influential frameworks due to their strong performance and transferability across downstream tasks.

More recently, Transformer-based architectures have emerged as competitive alternatives. The Audio Spectrogram Transformer (AST) \cite{gong2021ast} demonstrated that self-attention mechanisms can achieve state-of-the-art performance in audio classification, while subsequent works such as PaSST \cite{koutini2021efficient} introduced efficiency-oriented modifications to reduce the computational overhead associated with Transformer models. Additional research has explored lightweight architectures, knowledge distillation, and representation learning techniques aimed at improving recognition performance while reducing model complexity \cite{schmid2023efficient}.


\subsection{Edge AI for Audio Processing}\label{subsec:edge_audio}

The growing demand for real-time acoustic intelligence has stimulated increasing interest in deploying machine listening systems directly on edge devices. In contrast to cloud-based inference, edge computing enables lower communication latency, reduced bandwidth requirements, enhanced privacy preservation, and improved operational autonomy \cite{al2024computing,trigkas2026federated}.

Several studies have investigated the deployment of audio classification and sound event detection systems on embedded hardware platforms. Research in this area has predominantly focused on model compression, pruning, quantization, and TinyML techniques aimed at reducing computational requirements while maintaining acceptable predictive performance. Recent surveys on embedded audio intelligence identify energy efficiency, memory limitations, and computational constraints as major challenges for large-scale deployment of machine listening applications in real-world environments \cite{gairi2025environmental,vandendriessche2021environmental}.

A parallel research direction has investigated edge-based acoustic monitoring systems for smart cities, industrial environments, wildlife monitoring, and environmental sensing. These studies demonstrate the practical feasibility of deploying machine listening algorithms on resource-constrained platforms but typically evaluate only a single model architecture or deployment configuration \cite{antonini2018smart,nogueira2022sound}.

Consequently, although the literature has extensively explored how to reduce model complexity for edge deployment \cite{belatik2025optimization,singh2023edge}, relatively little attention has been devoted to understanding how changes in the underlying software ecosystem affect the execution characteristics of already deployed models \cite{khan2026edge}.

\subsection{Edge AI Deployment and Software Stack Evolution}
\label{subsec:edge_deployment}


Existing benchmarking studies have evaluated deployment-oriented metrics such as inference latency, throughput, memory footprint, energy consumption, and power efficiency \cite{qasaimeh2021benchmarking,hernandez2024optimizing,ratul2025accelerating}. Most works focus on comparing neural network architectures, numerical precision formats (FP32, FP16, and INT8), quantization methods, model compression techniques, or inference acceleration frameworks such as TensorRT \cite{vandendriessche2021environmental,gairi2025environmental,suchy2026review}. These studies consistently demonstrate that deployment efficiency cannot be inferred directly from predictive performance alone, highlighting the need to consider resource utilization and operational efficiency alongside recognition accuracy.

Modern Edge AI deployments, however, rely on complex software ecosystems comprising operating system components, hardware drivers, CUDA libraries, deep learning frameworks, compiler toolchains, and inference optimization engines. These software layers evolve continuously through successive releases that introduce optimized kernels, improved memory management strategies, enhanced operator implementations, and more efficient execution pipelines \cite{khan2026edge}. Despite their potential impact on deployment behavior, software stack versions are typically treated as fixed experimental conditions rather than as variables of interest. Consequently, benchmarking studies generally report results obtained under a single software configuration, making it difficult to assess the extent to which software evolution influences deployment performance.

This limitation is particularly relevant in machine listening. Although extensive research has investigated audio tagging architectures such as PANNs, AST, and PaSST \cite{Kong:2020:TASLP,gong2021ast,koutini2021efficient}, previous studies have primarily focused on recognition performance, model efficiency, and hardware-oriented deployment optimizations. To the best of our knowledge, no prior work has systematically analyzed how successive generations of the NVIDIA JetPack ecosystem affect the deployment characteristics of state-of-the-art audio tagging models under identical hardware conditions. In particular, it remains unclear whether software updates affect convolutional and Transformer-based architectures uniformly or whether certain architectural paradigms benefit disproportionately from improvements introduced in newer software releases.

Therefore, this work investigates software stack evolution as an independent factor in edge AI evaluation. By analyzing representative CNN- and Transformer-based audio tagging models on the same NVIDIA Jetson Orin Nano platform under JetPack~6.2 and JetPack~7.2, we provide new insights into how software ecosystem evolution influences latency, energy consumption, and resource utilization in embedded machine listening systems.

\section{Method}\label{sec:method}






\subsection{Evaluated Audio Tagging Architectures}\label{subsec:models}

Three widely adopted audio tagging architectures were selected to represent different design paradigms within modern machine listening systems.

\subsubsection{Pretrained Audio Neural Networks (PANNs)}

PANNs \cite{Kong:2020:TASLP} constitute one of the most influential CNN-based frameworks for audio tagging and audio pattern recognition. Trained on the large-scale AudioSet dataset, PANNs demonstrated that pretrained convolutional models can achieve strong generalization and transferability across a wide range of machine listening tasks. The framework encompasses multiple architectures with different depths, complexities, and feature extraction strategies, ranging from lightweight convolutional models to deeper hybrid architectures combining waveform- and spectrogram-based representations.

To capture this architectural diversity, our evaluation includes representative models from two major CNN families within the PANNs ecosystem. First, we consider the standard CNN architectures CNN6, CNN10, and CNN14, together with the Wavegram-Logmel-CNN14 (Wavegram-Logmel) variant, which combines learned waveform representations with log-mel spectrogram features and achieved state-of-the-art performance on AudioSet. Second, we evaluate the ResNet-based \cite{he2016deep} family through ResNet22, ResNet38, and ResNet54, which introduce residual connections to facilitate deeper feature extraction while preserving the overall convolutional design philosophy.

Due to their convolutional nature and broad adoption in audio tagging research, these architectures provide a representative benchmark for analyzing how software stack evolution affects CNN-based machine listening models. Furthermore, the diversity in depth, parameter count, and architectural design within the PANNs framework enables us to assess whether software-induced performance variations remain consistent across different convolutional paradigms.

\subsubsection{Audio Spectrogram Transformer (AST)}

The Audio Spectrogram Transformer (AST) \cite{gong2021ast} represents a major shift from convolution-based audio tagging architectures toward fully attention-driven models. Inspired by the Vision Transformer paradigm, AST converts log-mel spectrograms into sequences of overlapping spectrogram patches that are processed as tokens by a standard Transformer encoder. This design eliminates convolutional inductive biases and enables the model to capture long-range temporal and spectral relationships through self-attention mechanisms operating over the entire input sequence \cite{gong2021ast}.

As one of the first convolution-free architectures to achieve state-of-the-art performance on AudioSet and other audio classification benchmarks, AST established the viability of Transformer-based approaches for machine listening tasks \cite{gong2021ast}. From a deployment perspective, its attention-dominated computational profile differs substantially from that of CNN-based models, involving large matrix multiplications, memory-intensive attention operations, and execution patterns that heavily depend on optimized software kernels and runtime implementations.

Consequently, AST provides a representative Transformer baseline for investigating whether software stack evolution affects attention-based architectures differently from convolutional models, thereby enabling a direct comparison between two fundamentally distinct computational paradigms in edge audio intelligence systems.

\subsubsection{Patchout Audio Spectrogram Transformer (PaSST)}

PaSST (Patchout Audio Spectrogram Transformer) \cite{koutini2021efficient} extends the AST paradigm by introducing Patchout, a structured patch-dropping strategy that reduces the computational and memory requirements associated with Transformer-based audio processing while preserving, and in some cases improving, classification performance. The approach addresses one of the main limitations of audio Transformers, namely the quadratic growth of computational complexity with input length, by selectively discarding spectrogram patches during training and thereby reducing the amount of information processed by the attention mechanism.

By combining the representational power of self-attention with a design explicitly oriented toward computational efficiency, PaSST has become one of the most prominent Transformer-based architectures for large-scale audio tagging~\cite{primus2022improved,li2023fine}. Its efficiency-oriented design makes it particularly relevant for edge deployment scenarios, where memory footprint, inference latency, and energy consumption are critical constraints. 


Collectively, PANNs, AST, and PaSST provide a representative spectrum of contemporary audio tagging approaches, spanning convolution-based, Transformer-based, and efficiency-oriented Transformer architectures. This diversity enables a comprehensive assessment of whether the benefits introduced by successive software stack generations are architecture-dependent or remain consistent across fundamentally different computational paradigms.

\subsection{Software Stack Comparison}

The central variable investigated in this work is the software ecosystem used for model deployment. Two software environments were evaluated:


\begin{table}[t]
\centering
\caption{Software stack versions evaluated in this study.}
\label{tab:software_stack}
\begin{tabular}{lcc}
\hline
\textbf{Component} & \textbf{JetPack 6.2} & \textbf{JetPack 7.2} \\
\hline
L4T Release      & R36.4.7     & R39.2.0 \\
Ubuntu           & 22.04.5     & 24.04.4 \\
Kernel           & 5.15.148-tegra & 6.8.12-tegra \\
CUDA             & 12.6.68     & 13.2.75 \\
cuDNN            & 9.3.0       & 9.20.0 \\
cuBLAS           & 12.6.1.4    & 13.4.0.1 \\
TensorRT         & 10.3.0      & 10.16.2 \\
Python           & 3.10.12     & 3.12.3 \\
ONNX Runtime     & 1.20.0      & 1.24.0 \\
\hline
\end{tabular}
\end{table}

Both JetPack versions include different generations of CUDA libraries, TensorRT optimization engines, drivers, operating system components, and supporting runtime frameworks. Since these software layers directly influence model execution, memory management, kernel scheduling, and hardware acceleration, performance differences may emerge even when running identical models on the same hardware platform (see Table~\ref{tab:software_stack}).

The objective is not merely to compare absolute performance values between software versions but to determine whether software evolution affects all architectures uniformly or whether certain architectural paradigms benefit disproportionately from newer software optimizations.

\subsection{Inference Backends}

To capture the impact of software evolution across different execution paths, each model was evaluated using three deployment configurations available through ONNX Runtime.

\begin{itemize}
    \item \textbf{CPU Execution}: The first configuration performs inference exclusively on the ARM CPU of the target device. This scenario represents environments where GPU acceleration is unavailable or intentionally disabled due to energy or resource constraints.
    \item \textbf{CUDA Execution Provider}: The second configuration uses ONNX Runtime with the CUDA Execution Provider, enabling inference on the embedded NVIDIA GPU while relying on CUDA-based operator implementations.
    \item \textbf{TensorRT Execution Provider}: The third configuration employs ONNX Runtime integrated with TensorRT. In this mode, the runtime generates optimized inference engines that exploit hardware-specific optimizations, kernel fusion strategies, reduced precision execution, and Tensor Core acceleration whenever applicable.
\end{itemize}

Comparing these execution providers enables the analysis of how software stack evolution influences both generic GPU execution and highly optimized inference pipelines.

\subsection{Evaluation Metrics}

The evaluation focuses on deployment-oriented metrics commonly used in edge AI systems.

The following metrics were collected during experimentation:

\begin{itemize}
    \item Inference latency
    \item Energy consumption
    \item CPU/GPU utilization
    \item Memory usage
    \item Device CPU temperature
\end{itemize}

Latency characterizes real-time performance, while energy and CPU/GPU utilization measurements quantify deployment efficiency. Resource utilization and temperature measurements provide additional insight into the mechanisms underlying performance variations observed across software generations.

By jointly analyzing these metrics, the study aims to provide a comprehensive characterization of how software evolution influences edge deployment behavior for modern audio tagging architectures.

\section{Experimental Details}\label{sec:experimental}

All experiments were conducted on an NVIDIA Jetson Orin Nano developer platform running two software environments: JetPack 6.2 and JetPack 7.2. The hardware configuration, model weights, input data, and deployment pipeline were kept identical across all experiments to ensure that the software stack version remained the only independent variable under evaluation.


The selected audio tagging models were exported to ONNX format and executed using ONNX Runtime. For each model, inference was performed using three execution providers: CPU Execution Provider, CUDA Execution Provider, and TensorRT Execution Provider. This resulted in a total of eighteen deployment configurations corresponding to the combination of model architecture, JetPack version, and execution backend.


To strictly isolate the performance impact of software stack evolution from quantization artifacts, all models were exported and benchmarked using baseline single-precision floating-point (FP32). Maintaining FP32 precision eliminates confounding variables—such as quantization-induced accuracy degradation or mixed-precision fallback behaviors—thereby isolating how software updates alter fundamental execution pathways, layer fusion strategies, and hardware resource utilization across the underlying GPU architecture.

Consequently, the ONNX models preserve the numerical precision and predictive behavior of the official PyTorch implementations released by the original authors. This design choice ensures that any observed differences in latency, energy consumption, or resource utilization can be attributed to the execution backend and JetPack version rather than to model-level optimizations.

For every configuration, 1000 inference executions were performed using a dataset of 100 different audio clips (WAV format, 10\,s duration, 44.1\,kHz sampling rate) under identical operating conditions. During execution, system-level metrics were continuously monitored and recorded, including inference latency, energy consumption, CPU utilization, GPU utilization, memory usage, and device temperature.

To reduce the influence of transient effects and thermal accumulation, all experiments were executed in isolation, with a single inference process active at any time. After completing a measurement run, the device remained idle for 300\,s before the next configuration was evaluated. CPU, GPU, and memory clocks were pinned using \texttt{jetson\_clocks} throughout the evaluation to eliminate variability caused by dynamic frequency scaling. Thermal throttling indicators were also monitored during all experiments, and no throttling events were observed. Furthermore, all measurements were conducted under controlled environmental conditions (ambient temperature of 24\,$^{\circ}$C with active cooling).

To characterize the thermal conditions under which the experiments were performed, Table~\ref{tab:idle_temp} reports the idle temperature statistics measured for each JetPack version and power mode after the stabilization period. The results show that JetPack 7.2 consistently operated at lower idle temperatures than JetPack 6.2 across all power configurations. Nevertheless, temperature variability remained low in all cases, indicating stable operating conditions throughout the evaluation campaign.

\begin{table}[t]
\centering
\caption{Idle temperature statistics after the stabilization period.}
\label{tab:idle_temp}
\begin{tabular}{lcccc}
\hline
\textbf{JetPack} & \textbf{Mode} & \textbf{Mean ($^\circ$C)} & \textbf{SD} & \textbf{Range ($^\circ$C)} \\
\hline
6.2 & 15 W & 50.7 & 0.68 & 49.5--52.5 \\
6.2 & 25 W & 51.3 & 0.91 & 49.3--52.5 \\
6.2 & MAXN & 52.0 & 0.75 & 49.7--53.0 \\
\hline
7.2 & 15 W & 46.4 & 0.41 & 45.6--47.2 \\
7.2 & 25 W & 49.2 & 0.60 & 48.0--50.5 \\
7.2 & MAXN & 50.0 & 0.53 & 48.7--50.5 \\
\hline
\end{tabular}
\end{table}

This experimental protocol ensures a fair comparison between JetPack versions and enables the observed differences to be attributed primarily to software stack evolution rather than to variations in hardware state or execution conditions.

\section{Results}\label{sec:results}

\subsection{PANNs CNN Family}\label{subsec:panns_results}

\begin{figure}[t]
    \centering
    \includegraphics[width=1\linewidth]{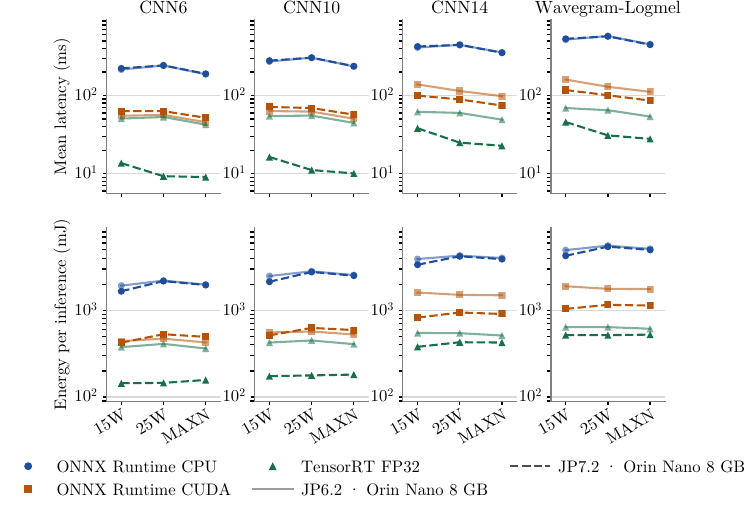}
    \caption{Mean inference latency (top row) and energy per inference (bottom row) for the CNN-based PANNs architectures (CNN6, CNN10, CNN14, and Wavegram-Logmel) deployed on the NVIDIA Jetson Orin Nano under JetPack~6.2 and JetPack~7.2. Results are reported for ONNX Runtime CPU, ONNX Runtime CUDA, and TensorRT FP32 execution providers across the 15\,W, 25\,W, and MAXN power modes.}
    \label{fig:panns_latency_energy}
\end{figure}

Figure~\ref{fig:panns_latency_energy} presents the latency and energy-per-inference results obtained for the CNN-based PANNs architectures under different power modes, inference backends, and JetPack versions.

The most significant result is the substantial impact of software stack evolution on TensorRT-based execution. While ONNX Runtime CPU and CUDA exhibit relatively modest differences between JetPack releases, TensorRT FP32 consistently benefits from JetPack~7.2 across all CNN architectures. The effect is particularly pronounced for the lightweight CNN6 and CNN10 models, whose latency decreases from approximately 50--60\,ms under JetPack~6.2 to below 10--15\,ms under JetPack~7.2, corresponding to speedups between 3.5$\times$ and 6$\times$ depending on the selected power mode. Although the improvement becomes less pronounced for larger architectures, CNN14 and Wavegram-Logmel still achieve latency reductions ranging from approximately 1.5$\times$ to 3$\times$. Since the hardware platform, model weights, and deployment configuration remain unchanged, these gains can be directly attributed to improvements introduced by the newer software stack.

Energy consumption follows a remarkably similar pattern. The magnitude of the energy reduction closely mirrors the latency improvements, indicating that JetPack~7.2 improves computational efficiency rather than simply increasing power draw to obtain faster execution. For CNN6 and CNN10, TensorRT energy consumption decreases from roughly 400--500\,mJ to around 150--200\,mJ per inference, representing reductions approaching a factor of three. CNN14 and Wavegram-Logmel also benefit from lower energy requirements, although the relative improvements are smaller, typically between 1.3$\times$ and 1.7$\times$. The strong correspondence between latency and energy trends suggests that the software stack improvements primarily reduce execution overheads and increase hardware utilization efficiency.

A second relevant observation is that the impact of JetPack~7.2 is clearly architecture dependent. The largest gains are consistently obtained for CNN6 and CNN10, whereas CNN14 and Wavegram-Logmel experience more moderate improvements. This behavior suggests that the optimizations introduced in the newer software stack disproportionately benefit workloads where runtime overheads constitute a significant fraction of the total execution time. As model complexity increases, a larger proportion of inference time becomes dominated by the neural network computation itself, reducing the relative influence of software-level optimizations.

Additional resource utilization metrics are reported in Appendix~\ref{app:resources} (Figure~\ref{fig:panns_resources}). The most striking change concerns GPU utilization during TensorRT execution. Under JetPack~6.2, GPU utilization typically remains below 40\%, reaching values as low as 10--15\% for CNN6 and CNN10. After upgrading to JetPack~7.2, utilization increases to approximately 95--100\% across almost all TensorRT configurations. This behavior strongly correlates with the latency and energy improvements observed in Figure~\ref{fig:panns_latency_energy}, suggesting that the newer software stack is able to exploit the available GPU resources far more effectively than its predecessor.

In contrast, CPU utilization remains essentially unchanged between software versions. ONNX Runtime CPU consistently occupies more than 85--90\% of the processor, while GPU-accelerated execution requires only a small fraction of CPU resources regardless of the JetPack release. Peak memory consumption also exhibits only moderate variations across software versions and does not show a consistent relationship with the observed performance improvements. Consequently, neither host-side CPU activity nor memory footprint appears sufficient to explain the substantial reductions in latency and energy consumption.

Overall, the results demonstrate that software stack evolution constitutes a major factor in the deployment efficiency of CNN-based audio tagging models. The transition from JetPack~6.2 to JetPack~7.2 yields substantial reductions in latency and energy consumption, particularly for TensorRT deployments, with speedups exceeding 4$\times$ for some lightweight architectures. The accompanying increase in GPU utilization suggests that a significant portion of these gains originates from improved exploitation of the underlying hardware resources rather than from changes in model execution semantics.

\subsection{PANNs ResNet family}\label{subsec:resnet_results}

\begin{figure}[t]
    \centering
    \includegraphics[width=1\linewidth]{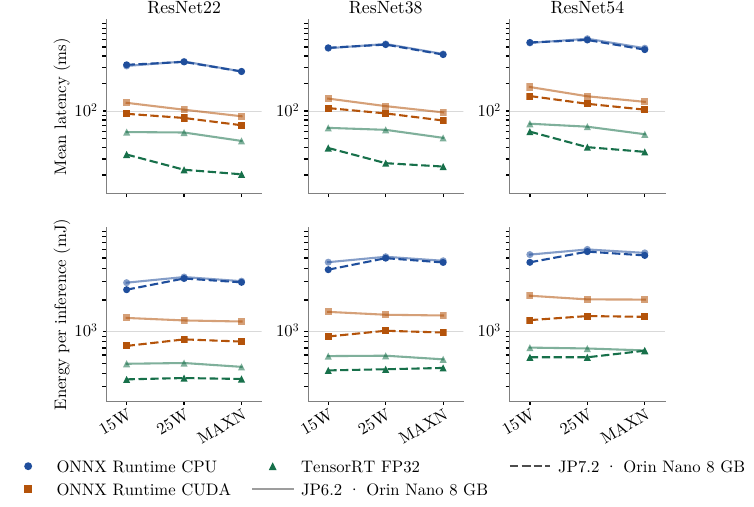}
    \caption{Mean inference latency (top row) and energy per inference (bottom row) for the ResNet-based PANNs architectures (ResNet22, ResNet38, and ResNet54) deployed on the NVIDIA Jetson Orin Nano under JetPack~6.2 and JetPack~7.2. Results are shown for ONNX Runtime CPU, ONNX Runtime CUDA, and TensorRT FP32 execution providers under the 15\,W, 25\,W, and MAXN operating modes.}
    \label{fig:resnet_latency_energy}
\end{figure}

Figure~\ref{fig:resnet_latency_energy} presents the latency and energy-per-inference results obtained for the ResNet-based PANNs architectures.

As observed for the CNN-based models (see Figure~\ref{fig:panns_latency_energy}), the software stack upgrade primarily benefits TensorRT-based execution. Across all ResNet variants, the transition from JetPack~6.2 to JetPack~7.2 systematically improves TensorRT performance, whereas the impact on ONNX Runtime CPU is negligible and only moderate improvements are observed for ONNX Runtime CUDA.

The latency reductions obtained with TensorRT are substantial but more moderate than those observed for the smallest CNN architectures. For ResNet22, latency decreases from approximately 65--70\,ms under JetPack~6.2 to around 30--40\,ms under JetPack~7.2, depending on the selected power mode. Similar reductions are observed for ResNet38, whose latency drops from roughly 70\,ms to around 30--40\,ms. ResNet54 also benefits from the software upgrade, although to a lesser extent, with latency decreasing from approximately 75--85\,ms to 45--60\,ms. In all cases, the reduction remains clearly below one order of magnitude, but still represents a considerable improvement in deployment efficiency without any modification to the model architecture or hardware platform.

Energy consumption follows the same overall pattern. TensorRT FP32 under JetPack~7.2 consistently provides the lowest energy-per-inference values across all ResNet architectures, with reductions that are generally on the order of 20--40\% relative to JetPack~6.2. ONNX Runtime CUDA also exhibits lower energy consumption under JetPack~7.2, while ONNX Runtime CPU remains largely unchanged between software versions.

Additional resource utilization metrics are reported in Appendix~\ref{app:resources} (Figure~\ref{fig:resnet_resources}). CPU utilization remains virtually identical across JetPack releases, with ONNX Runtime CPU saturating approximately one CPU core for all models. In contrast, GPU utilization increases substantially under JetPack~7.2. Under JetPack~6.2, TensorRT typically utilizes between approximately 15\% and 40\% of the GPU depending on model complexity and power mode, whereas under JetPack~7.2 utilization approaches 95--100\% across all ResNet architectures. This behavior closely mirrors the latency reductions observed in Figure~\ref{fig:resnet_latency_energy} and suggests that the newer software stack is able to exploit the GPU resources much more effectively.

Peak memory consumption scales with model complexity, increasing from roughly 4\,GB for ResNet22 to almost 6\,GB for ResNet54 when executed using TensorRT FP32. However, memory usage remains largely stable across JetPack versions, indicating that the observed performance gains cannot be attributed to substantial changes in memory allocation. Similarly, CPU temperature follows comparable trends for both software stacks and therefore does not appear to be a determining factor in the measured latency improvements.

Overall, the ResNet family confirms the trends previously observed for the CNN-based PANNs architectures. Software stack evolution primarily benefits GPU-accelerated inference, producing consistent reductions in latency and energy consumption while substantially increasing GPU utilization. Although the magnitude of the gains is smaller than that observed for the most lightweight CNN models (see Figure~\ref{fig:panns_latency_energy}), the results clearly demonstrate that deployment performance depends not only on model architecture and hardware capabilities but also on the maturity and optimization level of the underlying software ecosystem.

\subsection{Cross-Architecture Comparison}\label{subsec:cross_architecture}

\begin{figure}[t]
    \centering
    \includegraphics[width=1\linewidth]{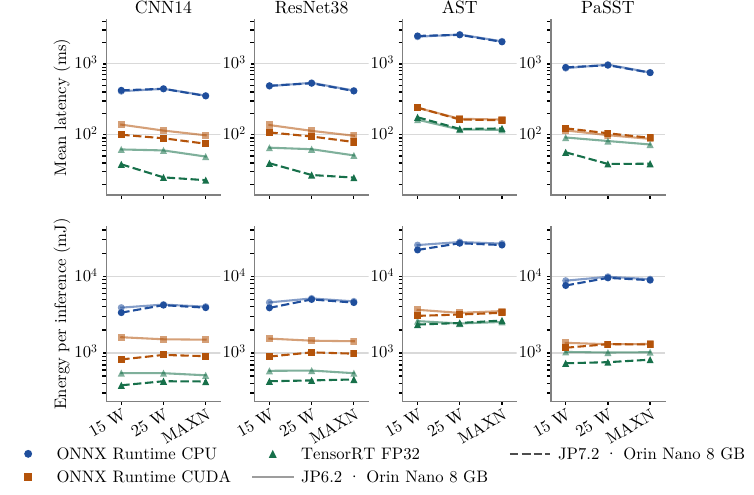}
    \caption{Mean inference latency (top row) and energy per inference (bottom row) for four representative audio tagging architectures: CNN14, ResNet38, AST, and PaSST. Measurements were obtained on the NVIDIA Jetson Orin Nano using JetPack~6.2 and JetPack~7.2, considering ONNX Runtime CPU, ONNX Runtime CUDA, and TensorRT FP32 execution providers under the 15\,W, 25\,W, and MAXN power configurations.}
    \label{fig:cross_latency_energy}
\end{figure}

To assess whether software stack evolution affects different architectural paradigms uniformly, Figure~\ref{fig:cross_latency_energy} compares four representative audio tagging models: CNN14, a widely adopted convolutional baseline; ResNet38, the highest-performing residual PANNs variant; AST, a pure Transformer architecture; and PaSST-S s16 SWA (PaSST), a Transformer specifically designed to improve efficiency.


The results reveal a clear architecture-dependent response to the software update. The convolutional models (CNN14 and ResNet38) obtain the largest benefits from the transition to JetPack~7.2 (specific details are discussed in previous sections).

AST exhibits a markedly different behavior. Although TensorRT execution also benefits from JetPack~7.2, the improvements are considerably smaller than those observed for the convolutional models. Latency decreases from approximately 170--180\,ms to around 120--160\,ms depending on the operating mode, while energy consumption remains within the same order of magnitude across JetPack releases. Consequently, AST remains substantially slower and more energy demanding than the CNN-based architectures under all deployment configurations considered.

PaSST presents an intermediate behavior. The model benefits from JetPack~7.2, particularly under TensorRT FP32, reducing latency from approximately 75--95\,ms to around 40--60\,ms while also lowering energy consumption. Although the magnitude of the gains is lower than that observed for CNN14 and ResNet38, it remains substantially higher than that obtained for AST. Moreover, PaSST consistently achieves lower latency and energy consumption than AST, demonstrating that Transformer-based models can remain competitive for edge deployment when architectural efficiency is explicitly considered during model design.

Additional deployment metrics reported in Appendix~\ref{app:resources} (Figure~\ref{fig:cross_resources}) help explain these differences. The most notable observation is that AST already operates at nearly 100\% GPU utilization under both JetPack releases and execution backends. Consequently, little opportunity exists for software-level optimizations to further increase hardware occupancy. In contrast, CNN14 and ResNet38 experience a dramatic increase in TensorRT GPU utilization (as seen in previous sections), rising from approximately 20--35\% under JetPack~6.2 to values approaching full utilization under JetPack~7.2. The substantial latency reductions observed for the convolutional models therefore appear to originate from improved exploitation of available GPU resources in the newer software stack.

Resource consumption also differs substantially across architectures. AST requires the largest energy budget and exhibits the highest operating temperatures, while CNN14 and ResNet38 achieve the lowest energy-per-inference values despite similar memory footprints. PaSST stands out as the most balanced Transformer model, combining significantly lower memory requirements than AST with considerably better latency and energy efficiency.

Overall, the results demonstrate that software stack evolution is not architecture agnostic. While convolutional models benefit substantially from the optimizations introduced in JetPack~7.2, the gains for Transformer-based architectures are considerably smaller. These findings indicate that deployment conclusions derived from a particular software generation may not transfer uniformly across model families and highlight the importance of considering software stack evolution when comparing competing audio tagging architectures for edge deployment.

\subsection{Accuracy--Latency Trade-off Across Architectures}
\label{subsec:accuracy_latency}

\begin{figure*}[t]
    \centering
    \begin{subfigure}[b]{0.49\textwidth}
        \centering
        \includegraphics[width=\linewidth]{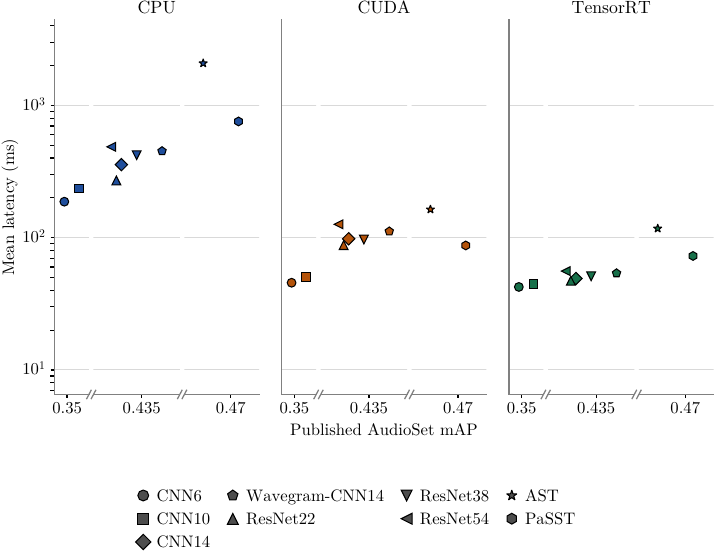}
        \caption{JetPack 6.2 (MAXN mode).}
        \label{fig:map_jp6}
    \end{subfigure}
    \hfill
    \begin{subfigure}[b]{0.49\textwidth}
        \centering
        \includegraphics[width=\linewidth]{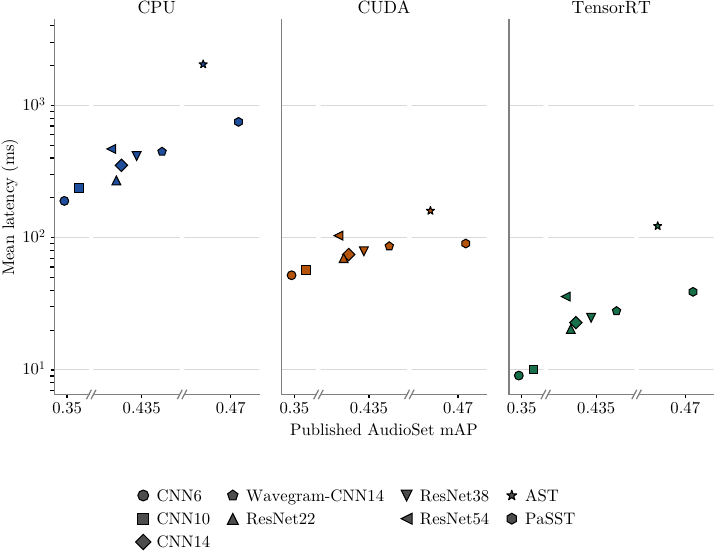}
        \caption{JetPack 7.2 (MAXN mode).}
        \label{fig:map_jp7}
    \end{subfigure}
    \caption{Accuracy--latency trade-off for all evaluated audio tagging models under different execution backends. The x-axis reports the published AudioSet mAP, while the y-axis shows the measured inference latency on the NVIDIA Jetson Orin Nano.}
    \label{fig:accuracy_latency}
\end{figure*}

While previous sections analyzed deployment characteristics independently, Figure~\ref{fig:accuracy_latency} evaluates the global accuracy--efficiency trade-off across all evaluated architectures by mapping published AudioSet performance against deployment latency on the Jetson Orin Nano.

Comparing the software stack generations reveals three key structural insights regarding the Pareto frontier:

\begin{itemize}
    \item \textbf{Non-linear Pareto efficiency:} Across all backends, model accuracy (mAP) does not scale linearly with deployment cost. Architectures yielding near-equivalent accuracy (e.g., CNN14 and ResNet variants around $0.42$--$0.44$ mAP) occupy radically different positions along the latency axis, confirming that predictive performance alone is a poor proxy for edge suitability.
    
    \item \textbf{Software-driven shift of the Pareto frontier:} Upgrading from JetPack~6.2 to JetPack~7.2 causes a structural reshaping of the efficiency frontier under TensorRT. Convolutional models experience a dramatic downward shift, establishing a new Pareto-optimal baseline that combines competitive mAP with minimal execution overhead.
    \item \textbf{Architecture-dependent frontier dynamics:} The evolution of the software stack alters the relative advantage between architectural paradigms. Under JetPack~6.2, the gap between CNNs and Transformers (such as AST and PaSST) is moderate. Under JetPack~7.2, CNNs and PaSST gain a disproportionate efficiency boost—moving PaSST into a highly favorable high-accuracy/low-latency sweet spot—while AST remains virtually static due to GPU saturation. 
\end{itemize}

Overall, these findings demonstrate that the accuracy--latency Pareto frontier is not a static hardware property, but a function of the software stack. Consequently, optimal model selection for edge deployment can shift across software releases even when model parameters and target hardware remain identical.

\section{Conclusion}\label{sec:conclusion}

This work investigated the impact of software stack evolution on the deployment characteristics of state-of-the-art audio tagging models on edge devices. Using an NVIDIA Jetson Orin Nano platform, we evaluated representative convolutional and Transformer-based architectures under two software generations, JetPack 6.2 and JetPack 7.2, while maintaining identical hardware conditions. The analysis considered multiple inference backends, power modes, and deployment metrics, including latency, energy consumption, resource utilization, and memory footprint.

The results demonstrate that software evolution can substantially influence deployment performance. Across most architectures, JetPack 7.2 consistently reduced inference latency and energy consumption, particularly when models were executed using TensorRT. More importantly, the observed improvements were strongly architecture dependent, revealing that software-level optimizations interact differently with the computational characteristics of each model family. In several cases, substantial deployment gains were achieved without modifying either the neural network architecture or the underlying hardware platform.

These findings indicate that deployment efficiency is determined not only by model design and hardware capabilities but also by the maturity and optimization level of the surrounding software ecosystem. Consequently, conclusions derived from benchmarking studies conducted under a single software configuration may not generalize across software generations, especially when comparing architectures with fundamentally different computational patterns.

More broadly, this work highlights the need to consider software stack evolution as an explicit experimental variable in Edge AI research. As embedded AI platforms continue to evolve, evaluating models exclusively from the perspective of predictive performance or hardware characteristics may provide an incomplete assessment of their real-world deployment potential. Reproducible deployment studies should therefore report software versions and execution environments with the same level of detail traditionally reserved for hardware specifications and model configurations.

Future work will extend this analysis to additional edge platforms, newer inference optimization frameworks, and lower-precision deployment pipelines (e.g., FP16 and INT8 quantization). In particular, we aim to investigate whether reduced numerical precision reveals new performance patterns and alters the efficiency gap between convolutional and Transformer-based models, while also dissecting the specific software-level mechanisms responsible for the architecture-dependent variations observed in this study.

\subsubsection{Acknowledgments.} 

This work was partially funded by the Valencian Institute for Business Competitiveness (IVACE). This work has been supported by Grants \mbox{PID2025-174060OB-C31 and PID2022-137048OA-C44 } funded by MICIU/AEI/10.13039/501100011033 and by ``ERDF/EU".

\clearpage

\section{Appendix}\label{sec:appendix}

\subsection{Appendix 1: Jetson Resources}\label{app:resources}

\begin{figure}[H]
    \centering
    \includegraphics[width=1\linewidth]{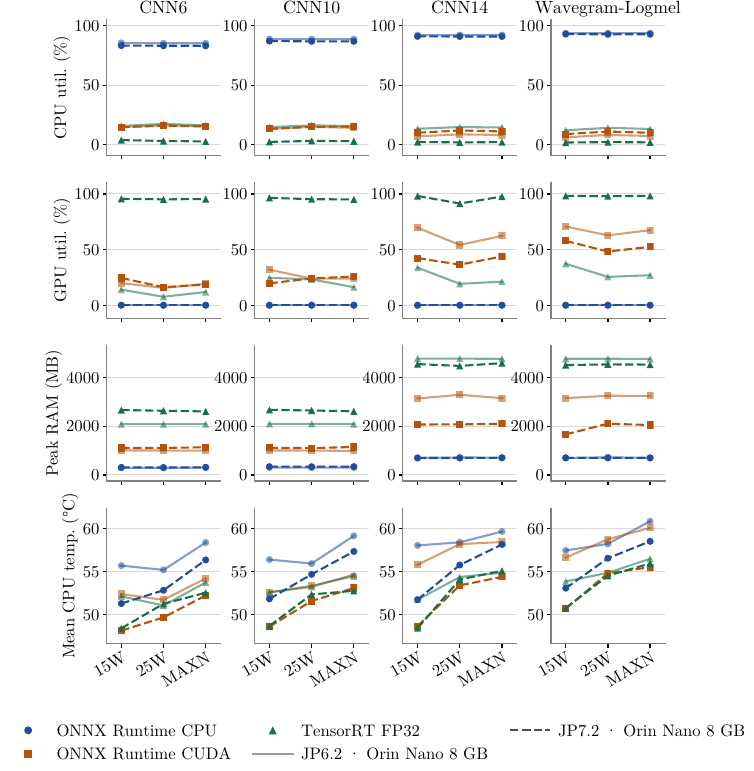}
    \caption{Resource utilization metrics for the CNN-based PANNs architectures (CNN6, CNN10, CNN14, and Wavegram-Logmel) under JetPack~6.2 and JetPack~7.2. From top to bottom: mean CPU utilization, mean GPU utilization, peak RAM usage, and mean CPU temperature. Results are reported for ONNX Runtime CPU, ONNX Runtime CUDA, and TensorRT FP32 across the 15\,W, 25\,W, and MAXN operating modes.}
    \label{fig:panns_resources}
\end{figure}

\begin{figure}[H]
    \centering
    \includegraphics[width=1\linewidth]{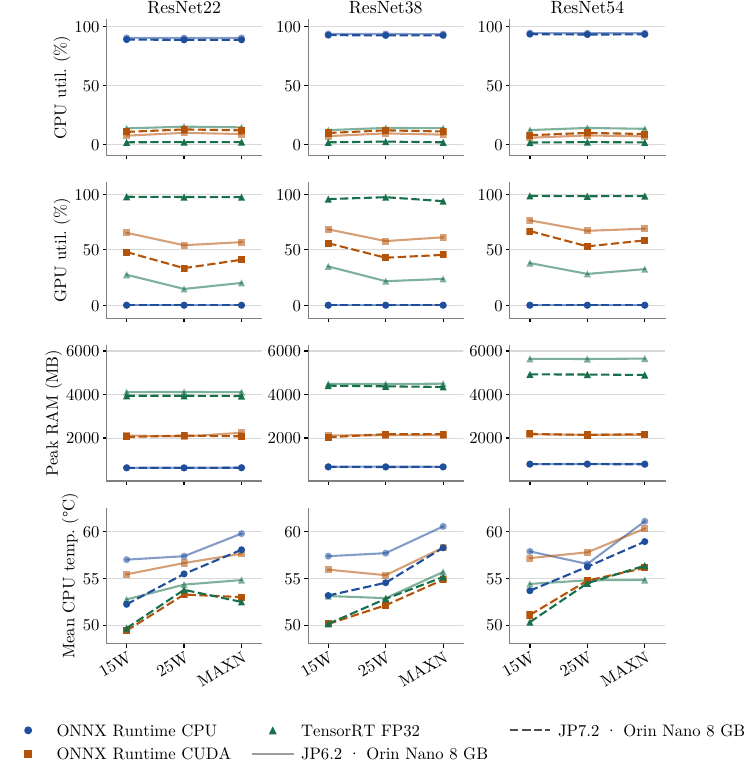}
    \caption{Resource utilization metrics for the ResNet-based PANNs architectures (ResNet22, ResNet38, and ResNet54) under JetPack~6.2 and JetPack~7.2. From top to bottom: mean CPU utilization, mean GPU utilization, peak RAM usage, and mean CPU temperature. Results are reported for ONNX Runtime CPU, ONNX Runtime CUDA, and TensorRT FP32 across the 15\,W, 25\,W, and MAXN operating modes.}
    \label{fig:resnet_resources}
\end{figure}

\begin{figure}[H]
    \centering
    \includegraphics[width=1\linewidth]{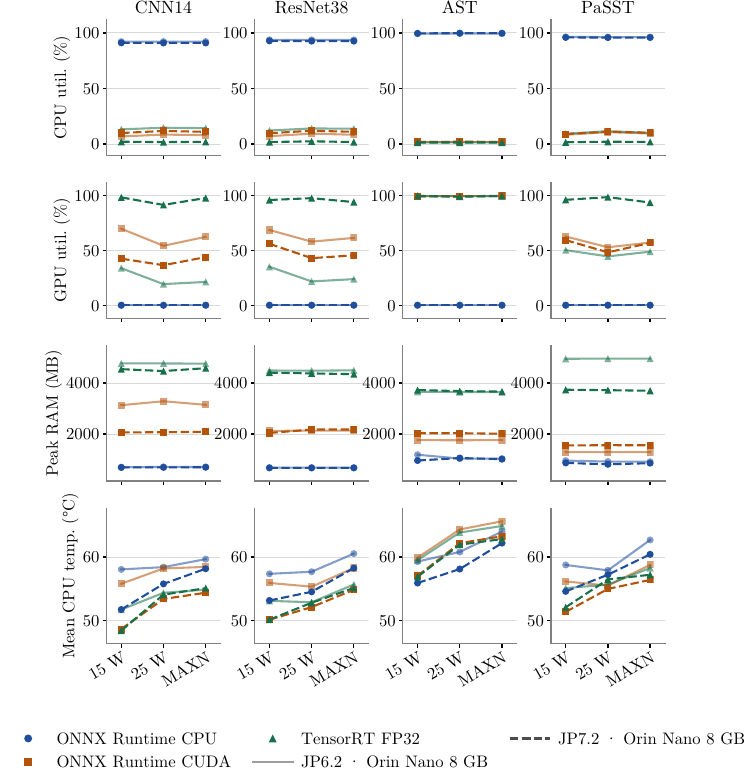}
    \caption{Resource utilization metrics for four representative audio tagging architectures (CNN14, ResNet38, AST, and PaSST) under JetPack~6.2 and JetPack~7.2. From top to bottom: mean CPU utilization, mean GPU utilization, peak RAM usage, and mean CPU temperature. Results are reported for ONNX Runtime CPU, ONNX Runtime CUDA, and TensorRT FP32 across the 15\,W, 25\,W, and MAXN operating modes.}
    \label{fig:cross_resources}
\end{figure}


%
%
%

\clearpage

\bibliographystyle{splncs03}
\bibliography{referencias}







\end{document}